\documentclass[10pt,twocolumn,letterpaper]{article}

\usepackage{iccv}              % To produce the CAMERA-READY version
\usepackage{lipsum}
\usepackage{comment}
\usepackage{graphicx}
\usepackage[accsupp]{axessibility}

\newcommand{\fname}{Text-to-character blendshapes~}
\newcommand{\sname}{T2Bs~}
\usepackage{xcolor} % optional (for gray text)
\newcommand\blfootnote[1]{%
  \begingroup
  \renewcommand\thefootnote{}% remove marker
  \footnote{\footnotesize\color{gray}{#1}}%
  \addtocounter{footnote}{-1}% don't step the counter
  \endgroup
}

\definecolor{iccvblue}{rgb}{0.21,0.49,0.74}
\usepackage[pagebackref,breaklinks,colorlinks,allcolors=iccvblue]{hyperref}

\def\paperID{7066} % *** Enter the Paper ID here
\def\confName{ICCV}
\def\confYear{2025}

\title{T2Bs: Text-to-Character Blendshapes via Video Generation}

\author{
Jiahao Luo\textsuperscript{1} \quad
Chaoyang Wang\textsuperscript{2} \quad
Michael Vasilkovsky\textsuperscript{2} \quad
Vladislav Shakhrai\textsuperscript{2} \quad
Di Liu\textsuperscript{3} \\
Peiye Zhuang\textsuperscript{2} \quad
Sergey Tulyakov\textsuperscript{2} \quad
Peter Wonka\textsuperscript{4} \quad
Hsin-Ying Lee\textsuperscript{2} \quad
James Davis\textsuperscript{1} \quad
Jian Wang\textsuperscript{2} \\
\textsuperscript{1}University of California, Santa Cruz \quad
\textsuperscript{2}Snap Inc.\quad
\textsuperscript{3}Rutgers University\quad
\textsuperscript{4}KAUST\\
{\tt\small \{jluo53, davisje\}@ucsc.edu, jwang4@snapchat.com}
}

\begin{document}
% \maketitle
% \begin{figure*}[h]
%   \centering
%   \includegraphics[width=\textwidth]{figs/fig1.jpg}
%   \vspace{-15pt}
%   \caption{\fname (T2Bs) is capable of creating animatable blendshapes to synthesize diverse expressions of a virtual character generated solely from text prompts.}
% \end{figure*} 

\twocolumn[{%
\maketitle

% \firstpagefootnote{Preprint. Under review at ICCV 2025.}

\begin{center}
\vspace{-22pt}
  \includegraphics[width=\textwidth]{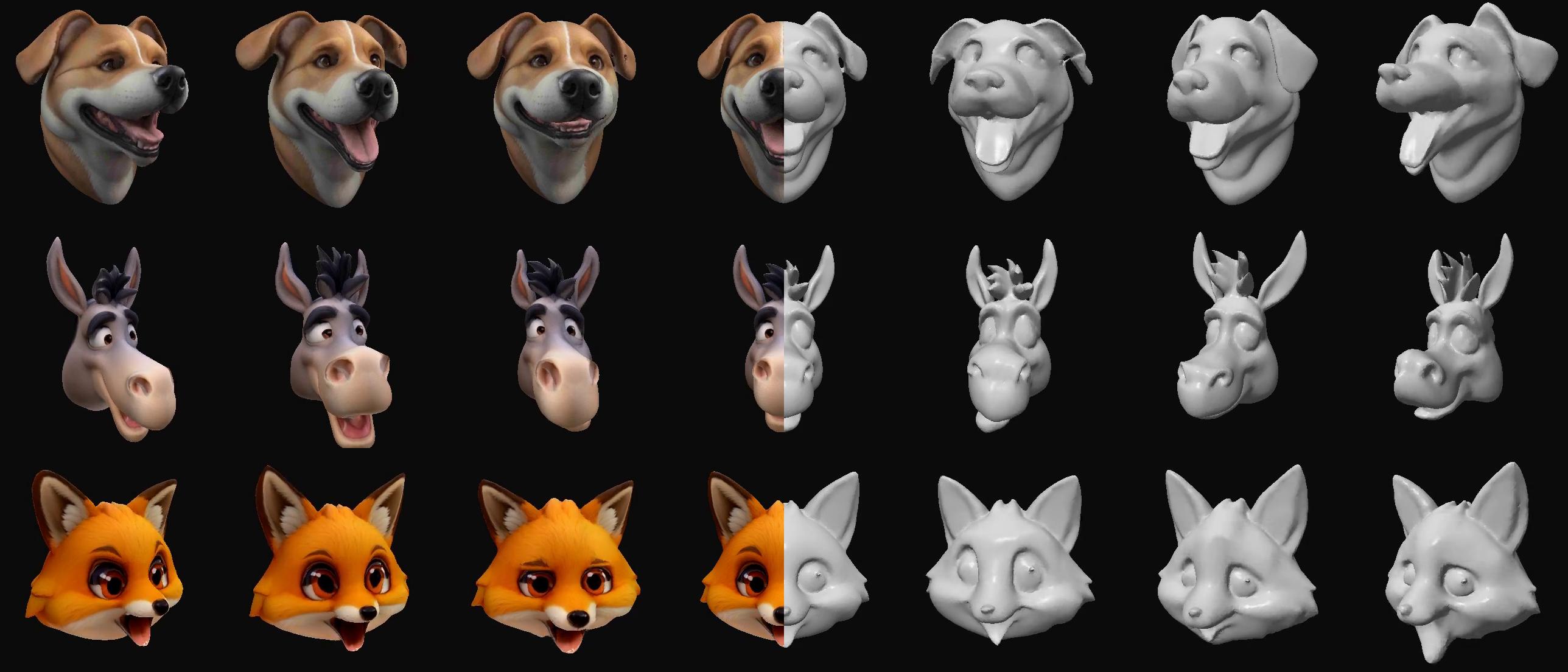}
  \vspace{-20pt}
  \captionof{figure}{\fname (T2Bs) is capable of creating animatable blendshapes to synthesize diverse expressions of a virtual character generated solely from text prompts.}
  \label{fig:main}
\end{center}
}]

\begin{abstract}
% We propose T2Bs, a new framework to create high-quality character head morphable models from text, combining static text-to-3D generation with video diffusion. Bridging the gap between these two methods is challenging: text-to-3D models produce detailed static geometry but cannot synthesize motion, while video diffusion models generate motion but face consistency issues like varying colors and geometric distortion. T2Bs uses deformable 3D Gaussian splatting to register static 3D models to video diffusion outputs, enabling the creation of a set of diverse, expressive motions with greater accuracy. By incorporating static geometry as a constraint and using a view-dependent deformation MLP, we reduce video artifacts and produce smooth, coherent, consistent results. T2Bs allows us to build 3DMMs that generate new, realistic expressions. Compared to existing 4D generation techniques, our method achieves superior results and creates expressive character head models that can be animated. \href{https://snap-research.github.io/T2Bs/}{Project Page}.
We present T2Bs, a framework for generating high-quality, animatable character head morphable models from text by combining static text-to-3D generation with video diffusion. Text-to-3D models produce detailed static geometry but lack motion synthesis, while video diffusion models generate motion with temporal and multi-view geometric inconsistencies. T2Bs bridges this gap by leveraging deformable 3D Gaussian splatting to align static 3D assets with video outputs. By constraining motion with static geometry and employing a view-dependent deformation MLP, T2Bs (i) outperforms existing 4D generation methods in accuracy and expressiveness while reducing video artifacts and view inconsistencies, and (ii) reconstructs smooth, coherent, fully registered 3D geometries designed to scale for building morphable models with diverse, realistic facial motions. This enables synthesizing expressive, animatable character heads that surpass current 4D generation techniques. Project Page: \href{https://snap-research.github.io/T2Bs/}{https://snap-research.github.io/T2Bs/}
\end{abstract}  
\vspace{-10pt}
\blfootnote{This work was done while Jiahao was an intern at Snap Inc.}
\section{Introduction}
\label{sec:intro}

The creation of animatable 3D virtual character head avatars \cite{zhou2024headstudio, han2023headsculpt, sang2022agileavatar, men20243dtoonify, mendiratta2023avatarstudio, wang2024headevolver} has become increasingly important due to their wide-ranging applications in social media, gaming, and entertainment. These avatars enable expressive, engaging, and highly personalized digital representations, enhancing both creative storytelling and interactive experiences. 3D human face models \cite{blanz2023morphable, egger20203d} have been extensively studied~\cite{flame, bfm09, cao2013facewarehouse, luo2022much} and widely applied in realistic~\cite{gaussianavatars, flashavatar, splattingavatar, gaussianhead, luo2025splatface} and virtual character~\cite{zhou2024headstudio, han2023headsculpt, mendiratta2023avatarstudio, wang2024headevolver} animation. However, the development of virtual character head models that deviate significantly from the human head distribution, such as animals, remains relatively underexplored. The unique challenges in this domain arise from the complexity of anatomy, the vast diversity of species, and the stylization often required for cartoon-like representations.
% Existing techniques~\cite{zuffi20173d} struggle to address these complexities effectively. 
Creating such virtual character models remains a labor-intensive process, typically requiring skilled artists and significant manual effort \cite{li2010example}. 

Recent advancements in text-to-3D generation, particularly those using diffusion models~\cite{dreamfusion, hifa, prolificdreamer, fantasia3d, dreamgaussian, gaussiandreamer}, have shown remarkable progress in creating high-quality static virtual character, including meshes and Gaussian representations~\cite{dmtet, FlexiCubes, liu2025lucaslayereduniversalcodec, 3dgs, liu2023deformer}. These methods are highly effective in generating detailed and realistic 3D geometry. However, their primary limitation lies in their inability to generate \textbf{diverse motion dynamics} that encompass the full range of movements a virtual character could exhibit.

% Start New Text
Video diffusion models, including image-to-video frameworks~\cite{cogvideox, snapvideo, SV3D, sora}, have shown promise in generating motion information. Several recent methods propose solutions on how to employ video models to generate dynamic 3D objects~\cite{dreamgaussian4d, sv4d, wu2024sc4d, li2024dreammesh4d, ren2024l4gm} combining techniques from video generation and 3D reconstruction. There are two main drawbacks we would like to improve upon. 
First, the methods have inherent limitations for generalization (model trained with limited 4D data) or efficiency (e.g., using score distillation sampling).
Second, the methods only tackle the problem of generating a single animation. This is not sufficient to obtain a 3D animatable model.

To address the first limitation, we propose a new framework that builds on recent 4D video generation techniques~\cite{wu2024cat4d, 4realVideo}. 4D video methods directly generate a multi-view video in the form of a grid of images. This enables a cleaner separation of motion generation and 3D reconstruction techniques and, ultimately, much higher visual quality.
However, 4D video still faces certain challenges, such as color inconsistencies, geometric distortions, and limited control over viewpoints and fine-grained details. 
% These issues can hinder their effectiveness in creating high-quality, expressive avatars.
% Even though we can build on deformable Gaussian splatting methods~\cite{4dgs,vidu4d,mosca}, these challenges are unique to the reconstruction from generated 4D videos.
To address the 3D inconsistency issues inherent in existing 4D generation methods, we introduce View-Conditioned Deformable Gaussian Splatting (VCDGS), which maintains structural coherence across views.

To address the second limitation, we use our framework multiple times with different text prompts to generate a larger variety of motions for a 3D character. Each initial text prompt results in one 4D video with a corresponding 3D mesh for each frame.
Using 3D meshes in a variety of poses, derived from many video sequences, we construct a blendshape model for the generated character that can be controlled to fit a variety of facial expressions. We evaluate our model and demonstrate its applicability in expression retargeting.

Our key contributions are as follows:
\begin{itemize}
    \item We introduce a scalable pipeline via multi-view video diffusion that overcomes data limitations, allowing the creation of blendshape models for a wide range of virtual characters from text prompts.
    \item We propose View-Conditioned Deformable Gaussian Splatting (VCDGS) to address the 3D inconsistency issue in existing 4D video methods, ensuring structurally coherent and high-quality deformations.
    \item We evaluate the expressiveness of our generated blendshape models and the potential for downstream animation and retargeting applications.
\end{itemize}

% Project page: \href{https://example.com/t2bs}{T2Bs}.

% \begin{itemize}
%     \item To the best of our knowledge, we are the first to tackle data limitations in developing morphable models for a wider range of virtual characters, introducing a scalable pipeline via multi-view video diffusion that creates blendshape models from text prompts.
%     \item We propose View-Conditioned Deformable Gaussian Splatting (VCDGS) to address the 3D inconsistency issue in existing 4D generation methods, ensuring structurally coherent and high-quality deformations.
%     \item We showcase the expressiveness of our generated blendshape model and its potential for downstream animation and retargeting applications.
% \end{itemize}

% These approaches demonstrate strong capabilities in animating 3D-aware geometry structures and natural motion patterns, offering potential solutions to the challenges posed by limited datasets to build virtual character head models. 
% Inspired by , we establish a connection between text-to-3D generation and video diffusion models to synthesize virtual characters with diverse deformations, enabling the construction of a head model that fully captures the range of movements a virtual character can exhibit. 
% To achieve this, we provide pre-defined text prompts and leverage the highly creative capacity of video diffusion models to generate expression-rich videos. 

% \vspace{-10pt}
\section{Related Works}

% \subsection{3D Generation}
\noindent\textbf{3D Generation:}
Text/image-to-3D generation has advanced significantly with diffusion models~\cite{ldm,DDPM}. DreamFusion~\cite{dreamfusion} pioneered high-quality text-to-3D generation by introducing Score Distillation Sampling (SDS) to leverage pretrained text-image diffusion models. Subsequent works~\cite{hifa, prolificdreamer, fantasia3d, dreamgaussian, gaussiandreamer, textmesh, luciddreamer, magic3d, makeit3d, gsgen, magic123} enhanced DreamFusion by refining SDS variants and incorporating diverse 3D representations~\cite{dmtet, HashEncoding, FlexiCubes, liu2025lucaslayereduniversalcodec, 3dgs, liu2023deformer, liu2023lepard} beyond NeRF~\cite{NeRF}. However, these methods remain computationally expensive due to iterative optimization during inference.
To accelerate 3D generation, recent approaches adopt a two-stage pipeline: (1) training image/video diffusion models to generate multi-view images~\cite{zero123, zero123pp, instant3d, SV3D, grm, mvdream, cat3d}, followed by (2) feedforward reconstruction methods~\cite{lrm, TriplaneGaussian, instantmesh} that rapidly synthesize 3D assets. Further, single-stage methods~\cite{ShapE, PointE, Rodin, RodinHD, MeshDiffusion, SDFusion, PrimDiffusion, SingleStageDiffusionNeRF, Michelangelo, NeuralWaveletDomainDiffusion, DiffRF} train diffusion models to directly generate explicit 3D representations.
%
% However, the generated 3D characters are static in nature and cannot be animated.
In this work, we build on existing 3D generation methods to create high-quality static 3D characters, while addressing an orthogonal problem: generating blendshapes for character morphing.

% Since our work on blendshape generation from video models is orthogonal to 3D generation, and as we do not impose any constraints on the input mesh, we utilize a proprietary text-to-mesh generation pipeline for initialization in all experiments.

% \vspace{-15pt}

%\subsection{4D Generation}
\noindent\textbf{4D Generation:}
4D generation is an emerging field with diverse definitions. Some approaches focus on generating videos with camera controls~\cite{vd3d,motionctrl,cameractrl,directavideo,4DiM}, while others generate a space-time video grid~\cite{sv4d,vividzoo,4diffusion,4real,4realVideo}. Our work aligns most closely with research that directly generates 4D representations, such as deformable Gaussian splats. This line of work~\cite{4dfy,ayg,consistent4d,dreamgaussian4d,4dgen,animate124,mav3d} leverages priors from text-to-image~\cite{ldm,imagen}, text-to-multiview~\cite{zero123,mvdream}, and text-to-video~\cite{animatediff,imagenvideo,videocrafter} models. However, most prior works~\cite{4dfy,ayg,consistent4d,dreamgaussian4d,animate124,mav3d, wu2024sc4d, li2024dreammesh4d, ren2024l4gm} generate independent 4D objects or scenes without constructing an animatable model that is capable of interactively generating new expressions or postures. The most related works to ours~\cite{gavatar,cap4d} use generative priors to create animatable human avatars, with morphable models of the human body~\cite{smpl} or face~\cite{flame} pre-trained on real human motion capture data. In contrast, our work is the first to learn a morphable model solely from generated data.

% NOT READY TODO!!!!!!!!! ADD more baselines

% DreamGaussian4D \cite{dreamgaussian4d} enables 4D generation from a single reference image.
% It first reconstructs a static 3D Gaussian Splat \cite{3dgs} using an enhanced version of DreamGaussian [cite] or other image-to-3D methods such as LRM \cite{lrm}. A video model \cite{stablevideodiffusion} is then used to synthesize a video from the input image. 4D generation is achieved by optimizing a deformation network on top of the static Gaussian Splat, supervised by thegenerated video and a multiview SDS loss based on Zero-123 XL \cite{zero123pp}. In addition,the method extracts a mesh sequence and refines the texture using an image-to-video  diffusion model.

% \vspace{-15pt}
%\subsection{Human Head Avatar}
\noindent\textbf{Human Head Avatars:}
Recent methods for creating head avatars utilize monocular or multi-view video inputs to synthesize new expressions. Among these, GaussianAvatars~\cite{gaussianavatars} rigs 3D Gaussians to the FLAME~\cite{flame} face tracking framework by anchoring them to the triangular facets of the mesh. Similarly, SplattingAvatar~\cite{splattingavatar} integrates 3D Gaussians into mesh models and predicts displacements along the normal direction. FlashAvatar~\cite{flashavatar} defines Gaussians in a uniform FLAME UV space and directly predicts per-Gaussian deformation from monocular video inputs. GaussianHead~\cite{gaussianhead} employs tri-plane representations and motion fields to simulate continuous geometric changes in heads, rendering rich textures, including skin and hair.
While these methods achieve impressive results, they rely heavily on face tracking and pre-existing human head models, limiting their applicability to scenarios involving characters that deviate significantly from human-like geometry. This constraint poses challenges for applications involving virtual characters or animal-based models, which fall outside the distribution of standard human datasets.

% {-5pt}
\section{Methods}
To build blendshape models from input text prompts, our pipeline consists of several stages. First, given a text prompt describing a virtual character, we use an off-the-shelf text-to-3D generator to create a textured 3D mesh. Next, a video diffusion model animates the 3D assets, generating a set of videos based on various text prompts describing different expressions. These text prompts are automatically generated using a prompt template, which substitutes the character description and the corresponding motion. In the subsequent capture stage, we employ a combination of advanced multi-view video generation models and robust reconstruction algorithms to capture the deformations of the 3D mesh from the generated videos. Finally, blending bases are computed from the captured deformed shapes. The pipeline is illustrated in Fig.~\ref{fig:framework}. The following sections provide a detailed explanation of each stage of the pipeline.

% {-5pt}
\subsection{Video Generation of Character Animations}
\label{sec:video_generation}
\subsubsection{Obtaining Diverse Expressions}
Starting with a textured 3D mesh obtained from a 3D generator, we collect a dataset of videos showcasing different expressions of the same character. To achieve this, we first create a pool of text prompts describing various expressions, such as `blinking eyes,' `frowning,' and `opening mouth.' Next, we render a frontal view of the 3D character, which serves as the conditional image input to a video diffusion model, which generates multiple videos corresponding to each prompt. To ensure the generated videos maintain a static camera pose and that the subject remains within the frame, we append ``static camera, the camera is holding still'' to the input prompts. These videos are image based only, often contain more than one animated component, and are not necessarily geometrically consistent, but they do contain a diverse range of expressions.

% \begin{figure*}[t]
%   \centering
%   \includegraphics[width=\textwidth]{figs/text2video_v2.pdf}
%   {-20pt}
%   \caption{
%   An illustration of generating 3D assets and multi-view videos from text prompts. First, we create a static 3D mesh using an off-the-shelf text-to-3D generator. Next, we render a freeze-time video with the camera moving horizontally in a circular path from left to right. From this, we select the frame corresponding to the frontal view and use an augmented prompt with predefined expressions to generate a fixed-view video. Finally, we apply a 4D video generation method to produce a set of multi-view videos.}
%   {-15pt}
%   \label{fig:alg1}
%   % \Description{}
% \end{figure*}

\begin{figure*}[t]
\centering
\includegraphics[width=\textwidth]{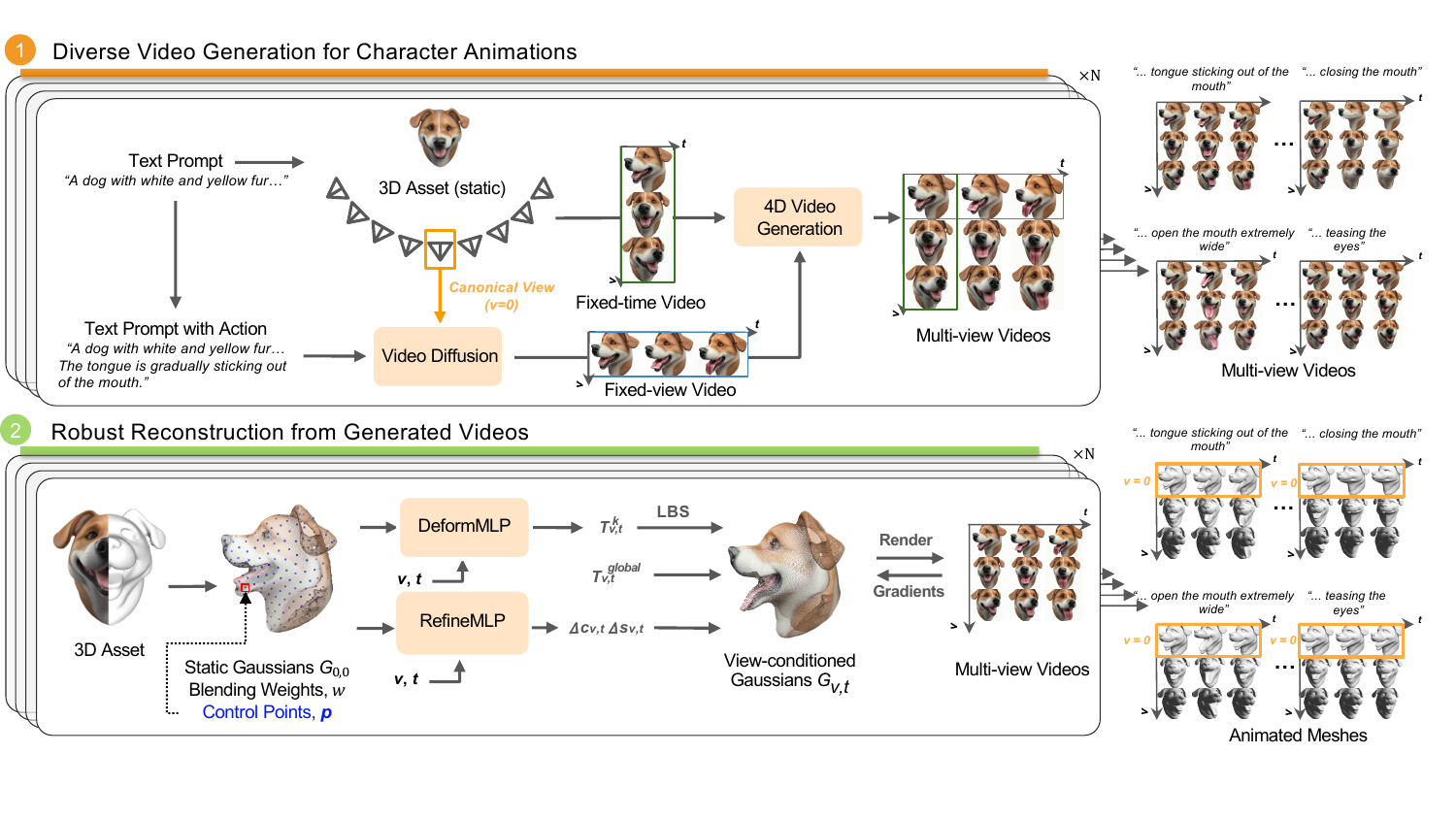}
\vspace{-40pt}
\caption{
\textbf{Overview of T2Bs.} In the first part, we illustrate the generation of multi-view videos from text prompts. A static 3D mesh is first created using an off-the-shelf text-to-3D generator~\cite{trellis3d}, followed by rendering a fixed-time video with the camera moving in a circular path. We define a canonical view (v=0) and an augmented prompt to generate a fixed-view video. A 4D video generation method is then applied to produce multi-view videos. In the second part, starting from a static 3D asset, we define static Gaussians $G_{0, 0}$, control points $p$ and blending weights $w$. During deformation, we predict view-dependent transformations of control points to model local non-rigid deformations, along with global transformations to capture overall pose changes. We interpolate Gaussian positions and orientations with Linear Blend Skinning (LBS), with rendering optimized through image-space loss minimization. After training, we extract a mesh for each frame, defined in the canonical view  (v=0). We repeat this process with multiple prompts, and build a blendshape model (shown in Fig. \ref{fig:main} 1st row) using hundreds of samples.
  }
\label{fig:framework}
\vspace{-5pt}
\end{figure*}

\subsubsection{Synchronized Multi-view Video Generation} 
The \emph{monocular} videos generated in the previous step provide supervision on how different expressions should appear in the frontal view. We refer to this as the fixed-view video. However, learning a 3D morphable model from frontal-view videos alone is ambiguous due to the under-constrained nature of the reconstruction problem. To address this, prior works have relied on either hand-crafted geometric or motion priors~\cite{vidu4d,som2024,mosca,wang2023flow}, which fail to produce high-quality results for complex motions such as tongue movements, or on score distillation sampling (SDS) loss~\cite{4real,dreamgaussian4d}, which is computationally expensive and typically requires several GPU hours to reconstruct even a short sequence. In this work, we explore the use of 4D video generation models capable of producing synchronized multi-view videos, offering more direct supervision for rendering expressions from multiple view points.

Specifically, the 4D video generation model produces a \emph{space-time frame grid}, ${\mathcal{I}_{v,t}}$, where $v$ and $t$ are indices representing the viewpoint and time, respectively. The model takes two input videos: the first is the fixed-view video showing the expression changes of the virtual character in the frontal view, \ie, $[\mathcal{I}_{0,0}, \mathcal{I}_{0,1}, \cdots, \mathcal{I}_{0,T}]$, assuming the frontal view index is 0. The second video is a fixed-time video showing viewpoint changes. This is rendered using the 3D mesh, with the camera moving circularly around the subject, \ie, $[\mathcal{I}_{0,0}, \mathcal{I}_{1,0}, \cdots, \mathcal{I}_{V,0}]$. Based on these inputs, the model generates the remaining frames in the grid, $\mathcal{I}_{v,t}$, $\forall v > 0, t > 0$. This process is repeated for the fixed-view video generated from each text prompt, to create a set of multi-view videos.

We experimented with two available 4D generation models, SV4D \cite{sv4d} and 4Real-Video~\cite{4realVideo}. We found that SV4D tends to produce blurrier results, particularly in the mouth and eye regions of the character. 
%We hypothesize that this is due to SV4D being trained on the animated Objaverse dataset, whose data distribution differs from our use case. 
In contrast, 4Real-Video consistently generates more plausible results. 
%This can be attributed to its foundation on video models trained with large-scale video datasets and its lightweight synchronization layers, which ensure spatial and temporal consistency across the generated frame grid while preserving the generalization capability of the original video model.

\noindent\textbf{Limitations of Multi-view Video Generation.} State-of-the-art 4D video model can generate visually consistent frames, however, directly applying vanilla 3D reconstruction yields noisy reconstruction due to several imperfections. First, the generated frames exhibit geometric \emph{multi-view inconsistency} and do not strictly adhere to epipolar constraints. Additionally, because current 4D models process only short durations of time in a single pass, they lack long-term context conditioning. As a result, these models inevitably produce \emph{inconsistent appearance} for regions that disappear and then reappear in the frame grid. To address these limitations, we propose a robust algorithm for 3D model reconstruction.

\subsection{Robust Reconstruction from Generated Videos}
\label{sec:reconstruction}
\subsubsection{Deformable 3D Gaussian Splats Representation} 
We represent the morphable model using deformable Gaussian splats, which offer high-quality rendering and fast performance. Specifically, we use static 3D Gaussian splats to represent the canonical model of the character. Each splat encodes its 3D position $\mathbf{x} \in \mathbb{R}^3$, orientation $\mathbf{q} \in SO(3)$, scale $\mathbf{s} \in \mathbb{R}^3$, and RGB color $\mathbf{c} \in \mathbb{R}^3$. To reduce reconstruction ambiguities, we omit higher-order spherical harmonics. Given the availability of the textured mesh, we initialize the canonical 3D Gaussian splats by cloning the vertex positions of the mesh, assigning the corresponding colors, and setting the opacity to 1. Subsequently, the scale and orientation of the splats are optimized by minimizing the rendering loss.

To deform the canonical 3D Gaussian splats, we adopted a linear blend skinning (LBS) formulation, {\it e.g.}~\cite{joshi2006learning,yang2021lasr,yang2022banmo}:
\begin{equation}
    \mathbf{x}_t = \sum_k w_k \mathbf{T}_t^k \mathbf{x},
    \label{eq:eq1}
\vspace{-10pt}
\end{equation}
where $\mathbf{x}_t$ is the transformed position at time $t$, $w_k$ represents the blending weight associated with the $k$-th deformation component, and $\mathbf{T}^k_t \in SE(3)$ denotes the corresponding rigid transformation. 
%The orientation of the Gaussian splats is transformed in a similar manner using LBS. 

We compute each control point \( \mathbf{p}_k \) and blending weight \( w_k \) in the rest pose, where \( \mathbf{p}_k \) is assigned via KNN and \( w_k \) via the Mahalanobis distance from each Gaussian to neighboring control points, using \( k = 2000 \) control points as detailed in the supplementary material. While \( w_k \) remains fixed, the per-control-point transformations \( \mathbf{T}_t^k \) are optimized by minimizing the following rendering loss:
\begin{equation}
   \sum_{v,t} \mathcal{L}_\text{huber}(\mathcal{I}_{v,t}-\mathcal{I}'_{v,t}) + \mathcal{L}_\text{LPIPS}(\mathcal{I}_{v,t}, \mathcal{I}'_{v,t}),
   \label{eq:loss}
\end{equation}
which combines an image-space Huber loss and a feature-space LPIPS loss~\cite{lpips}. This loss compares the frames generated by the video model, $\mathcal{I}_{v,t}$, with the frames rendered by the deformable Gaussian splats,  $\mathcal{I}'_{v,t}$.

% The deformation parameters $w_k$ and $\mathbf{T}^k$ are optimized by minimizing the following rendering loss:
% \begin{equation}
%    \sum_{v,t} \mathcal{L}_\text{huber}(\mathcal{I}_{v,t}-\mathcal{I}'_{v,t}) + \mathcal{L}_\text{LPIPS}(\mathcal{I}_{v,t}, \mathcal{I}'_{v,t}),
%    \label{eq:loss}
% \end{equation}
% which combines an image-space Huber loss and a feature-space LPIPS loss~\cite{lpips}. This loss compares the frames generated by the video model, $\mathcal{I}_{v,t}$, with the frames rendered by the deformable Gaussian splats,  $\mathcal{I}'_{v,t}$. 

\subsubsection{View-Conditioned Deformable Gaussian Splatting} Directly optimizing the loss in Eq.~(\ref{eq:loss}) yields unsatisfactory reconstruction due to imperfections in the frames generated by the video model. To mitigate these imperfections and improve quality, we propose {View-Conditioned} Deformable Gaussian Splatting (VCDGS).

\noindent\textbf{View-Dependent Deformation for Multi-view Inconsistency.} First, we propose modeling \emph{multi-view inconsistency} in the generated video frames as deformation. Specifically, we train a \emph{DeformMLP} that incorporates both the time $t$ and the view index $v$ as inputs to predict the rigid transformation for each LBS component. This approach ensures that the transformation depends on both $t$ and $v$, as described below:
\begin{equation}
    \mathbf{T}^k_{v,t} = \text{DeformMLP}(\mathbf{p}_k, v,t),
\end{equation}
where $\mathbf{p}_k \in \mathbb{R}^3$ denotes the position of the control point for the $k$-th deformation component. We note that the view-dependent transformations $\mathbf{T}^k_{v,t}$ are utilized only during training, when it is necessary to match the inconsistent multi-view video data. During testing, we designate $v=0$ as the canonical frame and use $\mathbf{T}^k_{0,t}$ to construct the final blendshape.

Additionally, we observe that directly training the DeformMLP can sometimes be unstable. To address this, we decompose the transformations into two parts: independent component-wise transformations predicted by DeformMLP and a shared rigid transformation representing the global $SE(3)$ motion of the object. Specifically,
\begin{equation}
    \mathbf{\tilde{T}}^k_{v,t} = \mathbf{T}_{v,t}^\text{global} \cdot \text{DeformMLP}(\mathbf{p}_k, v,t)
\end{equation}
During the initial stages of training, we first fit the global transformation $\mathbf{T}_{v,t}^\text{global}$ and then jointly optimize both the global transformation and the component-wise DeformMLP.

% \vspace{.3em}
\noindent\textbf{Gaussian Refinement for Appearance Inconsistency. } To capture appearance changes in the same region of the character across different frames generated by the video model, we introduce RefineMLP. This module predicts offsets for the non-positional properties of the Gaussian splats, including color, scale, and orientation, to better align with the generated videos. Specifically,
\begin{equation}
  \Delta \mathbf{c}_{v,t}, \Delta \mathbf{s}_{v,t}, \Delta \mathbf{q}_{v,t} = \text{RefineMLP}(\mathbf{x},  v, t)
\end{equation} 
Here, $\Delta \mathbf{c}_{v,t}$, $\Delta \mathbf{s}_{v,t}$, and $\Delta \mathbf{q}_{v,t}$ represent the refinement offsets for color, scale, and orientation, respectively. Similar to the view-dependent deformation, RefineMLP is used only during training and is not applied during inference nor in the construction of the final blendshape.

\subsection{T2Bs Blendshape Model and Retargeting}
 A set of deformed Gaussian splats are generated by the reconstruction process described in Sec.\ref{sec:reconstruction} for each video generated in Sec.\ref{sec:video_generation}. These splats are associated with mesh vertices so a 3D mesh representation for each frame of each video sequence is directly obtainable. These meshes are useful for rendering the existing video sequences, however they are not yet a blendshape model suitable for producing new animations.
 
 We  apply principal component analysis (PCA) to the per-frame geometry of all video sequences to construct a set of orthogonal blendshapes, eliminating the need for manually selecting blendshapes from specific frames based on expression prompts. This set of blendshapes is sufficiently expressive to model all observed deformations.

To produce new animations we retarget from human facial expressions to the domain of our blendshape model. We use semantically meaningful landmarks \cite{li2013realtime} that correspond to a subset of the standard 68 human facial landmarks. However, our virtual characters, particularly animal characters, have geometric distributions that deviate significantly from human faces, making off-the-shelf human landmark detection methods unsuitable.

To address this, we first render the static asset in a frontal view and use VLM models~\cite{ravi2024sam, liu2024grounding} to identify the left eye, right eye, and mouth regions. Next, we select eye and mouth landmarks along the edges of these segmentations at predefined angles. To map the selected 2D landmarks back to the static asset’s 3D space, we employ shadow mapping to locate the visible 3D points whose projections are closest to the identified 2D landmarks. Ultimately, we select 20 canonical landmarks that effectively cover the left eye, right eye, and mouth regions. We transfer motion from human landmarks to virtual character landmarks and then fit our orthogonal model to the deformed character landmarks. 

% Given human face tracking, we update the landmarks on the virtual character based on the relative positions of six landmarks on the left eye, six on the right eye, and eight on the mouth from the human facial landmarks. We then fit our blendshape model by optimizing its weights to minimize the L2 loss with as-rigid-as-possible (ARAP) regularization. Specifically, for the eye regions, we minimize the following objective:

% \begin{equation}
%     L = L_{\text{L2}} + \lambda_A L_{\text{ARAP}} + \lambda_D L_{\text{Depth}},
% \end{equation}
% where $L_{\text{Depth}}$ is an additional regularization term that prevents the eyeball geometry (segmented by the VLM model) from protruding in front of the eyelids (eye landmarks).

% We consider capture of VCGD per frame $G_{t}$

% % \begin{equation}
% %   x = \sum \left( w \cdot \text{trans} \cdot x_0 \right),
% % \end{equation}

% \begin{equation}
%   x_0 = x_{1,0} + B_e \cdot e,
% \end{equation}
% where $x_0$ is the position of the Gaussians. $x_{1, 0}$ is static asset position. $B_p$ is the base of pose dependent deformations and d is the coefficient. We perform PCA on the vertices of each part in a local coordinate frame, obtaining a vector of average deformations $m_{p,i}$ and the basis matrix Bp,i
% TT

\begin{figure*}[h!]
  \centering
  \includegraphics[width=\textwidth]{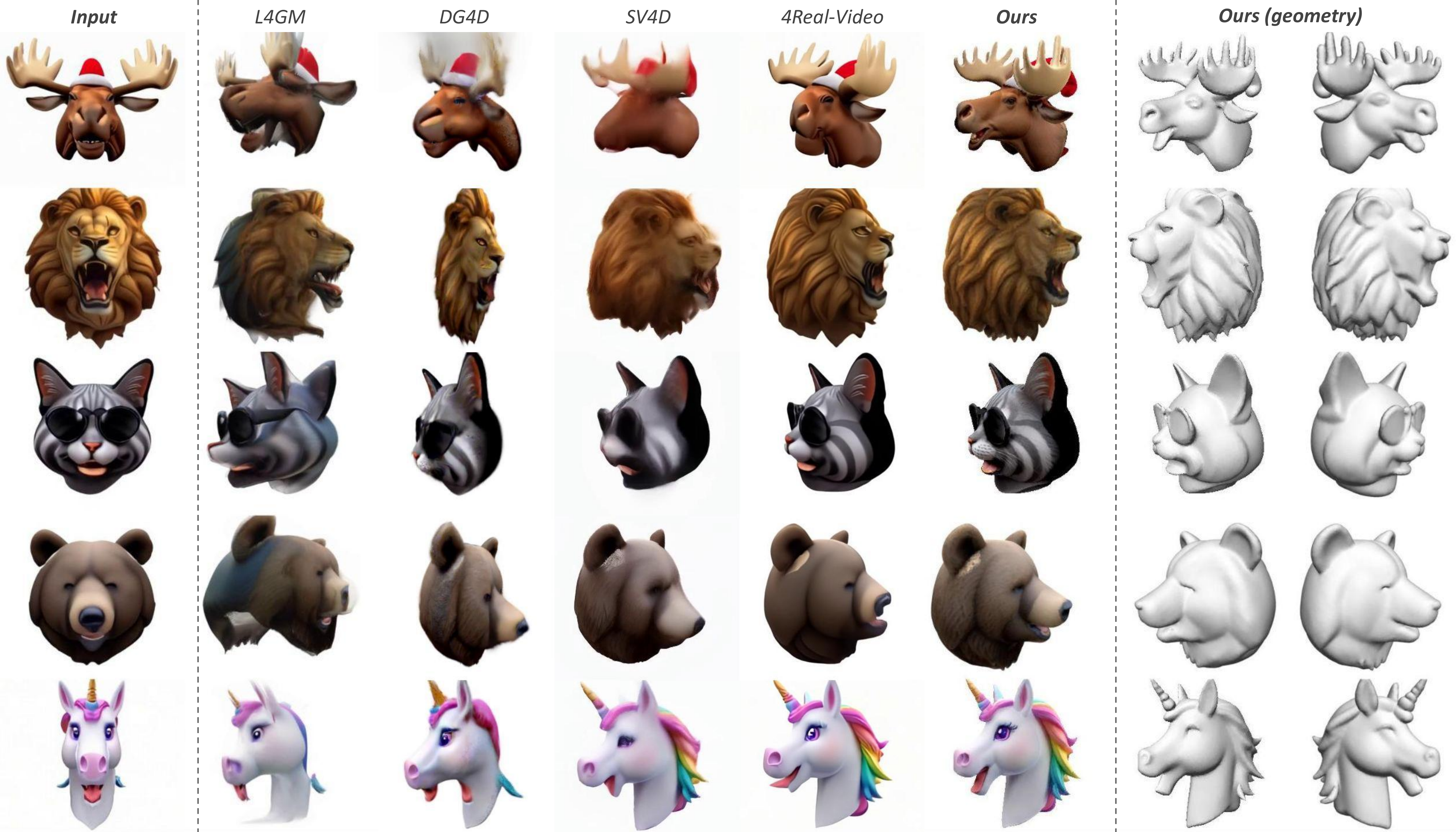}
  \caption{Qualitative comparison of 4D generation methods. We compare the 4D generation results of our method with L4GM \cite{ren2024l4gm}, DreamGaussian4D (DG4D) \cite{dreamgaussian4d}, SV4D \cite{sv4d}, and 4Real-Video \cite{4realVideo}. %Alongside a monocular video, DG4D uses our static Gaussian representation as input, while SV4D and 4Real utilize freeze-time renderings. We show viewpoints at ±60 degrees relative to the original view used to generate the monocular video. Among the compared methods, our approach produces the most visually appealing results.
  DG4D incorporates our accurate static Gaussian representation as input alongside a monocular video, while SV4D and 4Real rely on fixed-time renderings. Viewpoints are displayed at ±60 degrees relative to the original perspective used for generating the monocular video. Among all methods, our approach achieves the most visually consistent and appealing results. Note that, unlike other methods, ours also produces high-quality meshes. Video results are in the Supp. material.}
  \label{baseline}
  \vspace{-5pt}
\end{figure*}
\section{Experiments}

\subsection{Implementation Details}

% Given a static mesh with an average bounding box size of approximately 1.2, we render 15 images using a camera with a 120-degree field of view (FoV). The camera positions are defined in spherical coordinates with an elevation of 10 degrees, an azimuth linearly varying from -60 to 60 degrees, and a distance of 2.25 units from the mesh center (as shown at the top center of Fig. \ref{fig:alg1}). Each image is rendered at a resolution of $512\times512$, then cropped and resized to various dimensions based on the requirements of the image-to-video model and baseline methods. For baseline comparison, we use a cropped image size of $288\times512$.
% \vspace{-10pt}

When evaluating our method, we generate multi-view videos as described in Sec.~\ref{sec:video_generation}, rendering the 3D asset at a resolution of $512 \times 512$, followed by cropping as required by video models~\cite{cogvideox, 4realVideo}. Each video prompt combines a predefined expression prompt (from a set of 20, e.g., those shown on the right side of Fig.~\ref{fig:framework} with more details in Supp. material) with the original character prompt and camera motion description. For each concatenated prompt, we generate three sets of multi-view videos and manually filter out cases with unnatural expressions, retaining those that either align with the prompt or appear naturally plausible.
We train VCDGS for 60,000 iterations per set of multi-view videos. The entire process, including multi-view video generation and VCDGS optimization, takes approximately one hour on a single NVIDIA A100 GPU.

\subsection{VCDGS Evaluation}
% \subsubsection{Comparison Baselines}
\noindent\textbf{Comparison Baselines.}
We compare the proposed View-conditioned Deformable Gaussian Splatting (VCDGS) with recent state-of-the-art 4D generation methods: \emph{DreamGaussian4D}~\cite{dreamgaussian4d}, \emph{SV4D}~\cite{sv4d}, \emph{4Real-Video}~\cite{4realVideo} and \emph{L4GM}~\cite{ren2024l4gm}. For a fair comparison, we initialize DreamGaussian4D~\cite{dreamgaussian4d} with the same static Gaussian splats as our method before optimizing the deformation field.
For SV4D~\cite{sv4d}, we render a 21-frame, 360-degree video of the generated 3D asset and select 8 frames as input.
For 4Real-Video~\cite{4realVideo}, we follow the authors' implementation, rendering a 15-frame input video with azimuths from -60 to 60 degrees.
For L4GM~\cite{ren2024l4gm}, we adhere to the official implementation to generate Gaussian splats from the input fixed-view video.

\noindent\textbf{Qualitative Comparison.}
A qualitative comparison of 4D generation results between our method and baseline approaches is shown in Fig.~\ref{baseline}. Each method renders views at ±60 degrees from the original viewpoint used for monocular video generation. Reconstructing 4D assets from distant viewpoints remains highly challenging, even with access to the first frame.
Although initialized with the same static Gaussians as our method, DG4D produces distorted geometry. SV4D struggles with complex structures, such as moose antlers, resulting in blurry outputs. Similarly, 4Real-Video suffers from geometric distortions, limiting its performance. Our method overcomes these challenges by refining 4Real-Video’s output through a view-dependent design and integrating geometric cues from the static scene to resolve 3D inconsistencies. As a result, it achieves the most visually consistent and high-quality reconstructions among all baselines.
In addition to rendered images, our approach also generates a 3D mesh, shown on the right side of the figure.
% Note that our model takes the results of 4Real as input and improve beyond 4Real by utilize the geometry in static scene with the precision necessary to overcome the 3D inconsistency issue. Our method yields the most visually appealing results with respect to the input videos.

\noindent\textbf{Quantitative Evaluation.} Evaluating generated 3D/4D assets is challenging. Due to the limited number of generated assets, commonly used metrics like FID~\cite{bynagari2019gans} and FVD~\cite{fvd} are not statistically meaningful. While CLIP scores~\cite{clip} assess text-3D alignment, they lack flexibility for other criteria. Instead, we adopt the Elo rating~\cite{elo} from GPTEval3D~\cite{gpteval3d}, which leverages GPT-4o~\cite{gpt4} to compare 3D assets across multiple criteria and compute rankings based on pairwise comparisons.
Following the official implementation, we sampled 300 pairwise comparisons and report Elo ratings for three key criteria: \emph{3D plausibility}, \emph{texture details}, and \emph{text-asset alignment}, omitting criteria requiring surface normal rendering, as some baselines do not support it. As shown in Table~\ref{tab:gpt_eval}, our results are predominantly preferred by GPT-4o evaluations.

% We further evaluate our method on 10 artist-animated virtual characters that is closely aligned with our application domain from Objaverse-XL~\cite{deitke2023objaverse} (details in Supp. material). For each identity, we baked geometry sequences with a shared texture. We rendered (ambient=1.0) fixed-time, fixed-view videos as the pipeline input, and full sequences across all times and views as ground truth. We show the quantitative effect of proposed components in Sec.~\ref{ablation}.

% We evaluated the per-frame mesh reconstruction as shown in Tab.~\ref{tab:quantitative} (Middle), analyzing how geometry degrades when each proposed component is removed.

\noindent\textbf{User Study.}
We conducted a user study comparing DreamGaussian4D (DG4D)\cite{dreamgaussian4d}, SV4D\cite{sv4d}, 4Real-Video~\cite{4realVideo}, and our method. The study included 422 samples, each rated by 10 participants, yielding 4,220 ratings across four dimensions. \emph{Appearance} evaluates identity and detail preservation. \emph{Motion} measures the accuracy of motion transfer while avoiding scaling and rotation artifacts. \emph{Geometry} assesses distortions such as squishing or shape deformations. \emph{Overall} represents the plausibility of motion transfer while maintaining the appearance and geometry of the base asset. As shown in Table \ref{tab:preference_distribution}, VCDGS outperforms all baselines, including 4Real-Video, which it refines. This improvement stems from our view-dependent deformations, which resolve multi-view inconsistencies.

\begin{table}[t]
\centering
\renewcommand\tabcolsep{5pt}
\resizebox{0.9\linewidth}{!}{
\begin{tabular}{l|c|c|c|c}
    \toprule
    \textbf{Criteria (\%)} & \textbf{DG4D~\cite{dreamgaussian4d}} & \textbf{SV4D~\cite{sv4d}} & \textbf{4Real-Video~\cite{4realVideo}} & \textbf{Ours} \\
    \midrule
    Appearance & 0.0 & 0.2 & 13.0 & \textbf{86.8} \\
    Motion & 0.1 & 0.4 & 40.9 & \textbf{58.7} \\
    Geometry & 0.0 & 0.1 & 15.8 & \textbf{84.2} \\
    Overall & 0.1 & 0.1 & 15.3 & \textbf{84.6} \\
    \bottomrule
\end{tabular}
}
\vspace{-5pt}
\caption{User preference distribution across different methods.} % Our method received the highest overall preference with 84.6\%, significantly outperforming DG4D, SV4D, and 4Real-Video. \todo{decide whether to redo as the user pref}}
\label{tab:preference_distribution}
\end{table}

\begin{table}[t]
\centering
\renewcommand\tabcolsep{5pt}
\resizebox{\linewidth}{!}{
\begin{tabular}{l|c|c|c|c|c}
    \toprule
    \textbf{Criteria (\%)}  & \textbf{SV4D~\cite{sv4d}} & \textbf{L4GM~\cite{ren2024l4gm}}& \textbf{DG4D~\cite{dreamgaussian4d}} & \textbf{4Real-Video~\cite{4realVideo}} & \textbf{Ours} \\
    \midrule
    Text-Asset Align. &  818.5 & 729.9 & 1136.6 & 1075.2 & \textbf{1301.5} \\
    3D Plausibility& 799.4 & 734.1 & 1138.3 & 1126.7 & \textbf{1306.2} \\
    Texture Details & 664.1 & 624.5 & 1157.4 & 1151.4 & \textbf{1436.4} \\
    \bottomrule
\end{tabular}
}
\vspace{-5pt}
\caption{GPTEval3D~\cite{gpteval3d} ratings using GPT-4o, higher is better.} % Our method received the highest overall preference with 84.6\%, significantly outperforming DG4D, SV4D, and 4Real-Video. \todo{decide whether to redo as the user pref}}
\label{tab:gpt_eval}
\end{table}

% \begin{figure}[!t]
%     \centering
%     \begin{subfigure}{0.45\linewidth}
%         \centering
%         \includegraphics[width=\linewidth]{figs/radar_chart_trial.pdf}
%         \caption{User preference}
%         \label{fig:subfig1}
%     \end{subfigure}
%     \hfill
%     \begin{subfigure}{0.45\linewidth}
%         \centering
%         \includegraphics[width=\linewidth]{figs/radar_chart_trial.pdf}
%         \caption{GPTEval3D~\cite{gpteval3d} Elo score}
%         \label{fig:subfig2}
%     \end{subfigure}
%     \caption{Quantitative comparison of 4D generation methods.}
%     \label{fig:elo_rating}
% \end{figure}

\begin{figure}[!t]
  \centering
  \includegraphics[width=\linewidth]{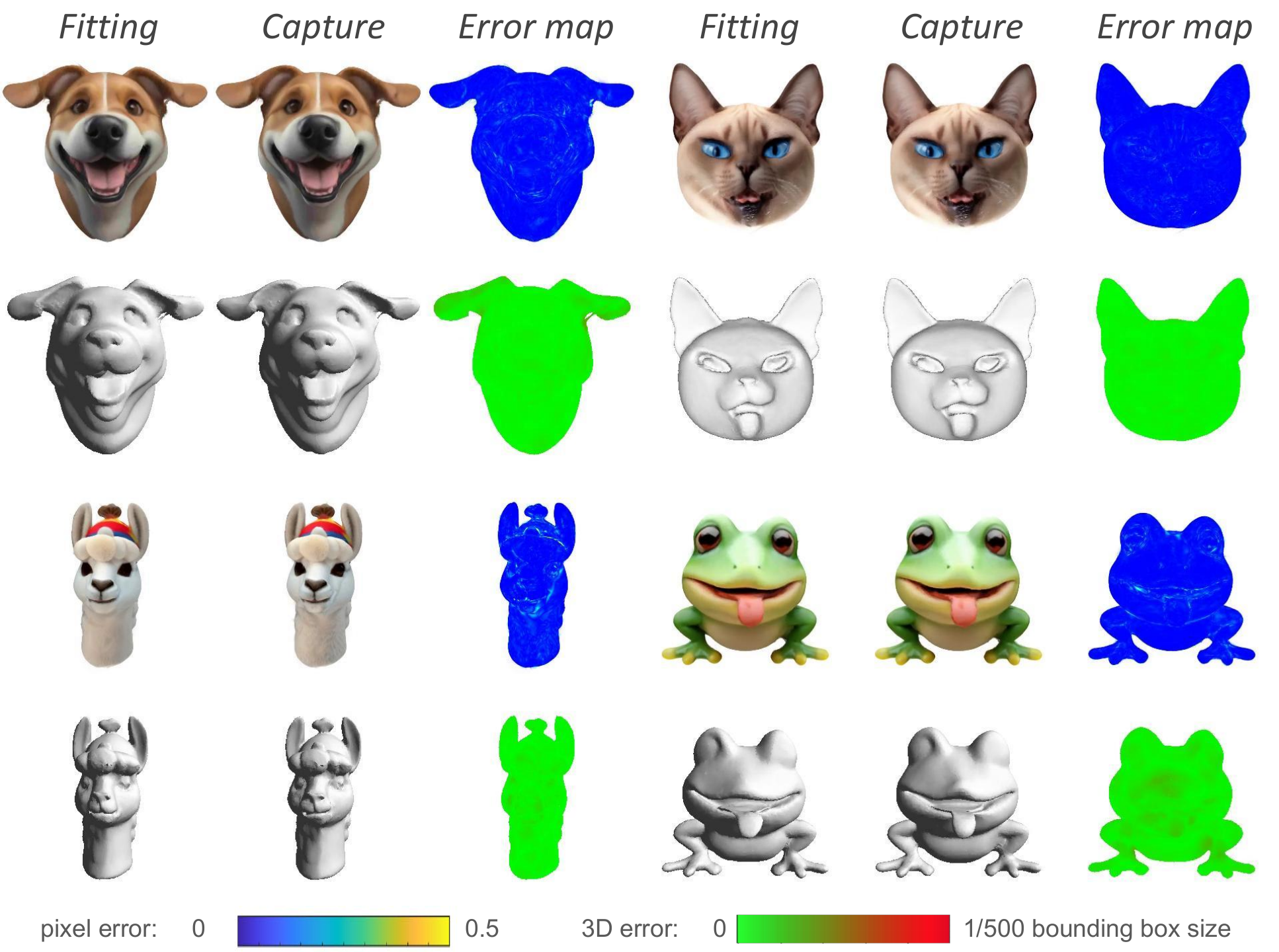}
  \caption{Fitting the learned blendshape model to held-out test set captures. The color scale from blue to yellow represents the RGB error, while the scale from green to red denotes the 3D point-to-point error.}
  \label{PCA_fit}
  \vspace{-10pt}
\end{figure}

\subsection{Blendshape Evaluation}
We evaluate the quality of blendshape models learned from VCDGS captures and demonstrate their applicability for expression transfer using a simple retargeting approach. Please refer to the project website for the video results.
% \vspace{-5pt}

\noindent\textbf{Model Expressiveness.}
A robust statistical expression model should effectively generalize to new data while remaining closely aligned with the specific object it represents. We fit T2Bs model with 100 blendshapes to geometry extracted from animations outside the model's training set by minimizing $\mathcal{L}_\text{geo} + \lambda_{rgb} L_{\text{rgb}}$,
where $L_{\text{geo}}$ is point-to-point euclidean distance and $L_{\text{rgb}}$ is the pixel-wise color distance when rendering the capture and the reconstruction with the same camera. We set $\lambda_{rgb} = 0.1$.

As shown in Fig.~\ref{PCA_fit}, the reconstructions closely align with the captures in both geometry and appearance. For each identity, the first column shows model fitting, the second displays a held-out video capture, and the third presents RGB and 3D error maps, comparing the capture and fitting results in terms of rendering and per-vertex differences. Our blendshape model generalizes effectively to new data, demonstrating its ability to faithfully reconstruct meshes of held-out expressions.

% \begin{figure}[!t]
%   \centering
%   \includegraphics[width=\linewidth]{figs/retargeting.pdf}
%   \caption{Sample expression retargeting from human to virtual characters. Top two rows: FLAME expressions and their retargeted counterparts on virtual characters, rendered as geometry. Bottom two rows: Real-world captures and corresponding retargeted expressions, rendered with texture.}
%   \label{sync_human}
% \end{figure}
\begin{figure}[!t]
  \centering
  \includegraphics[width=\linewidth]{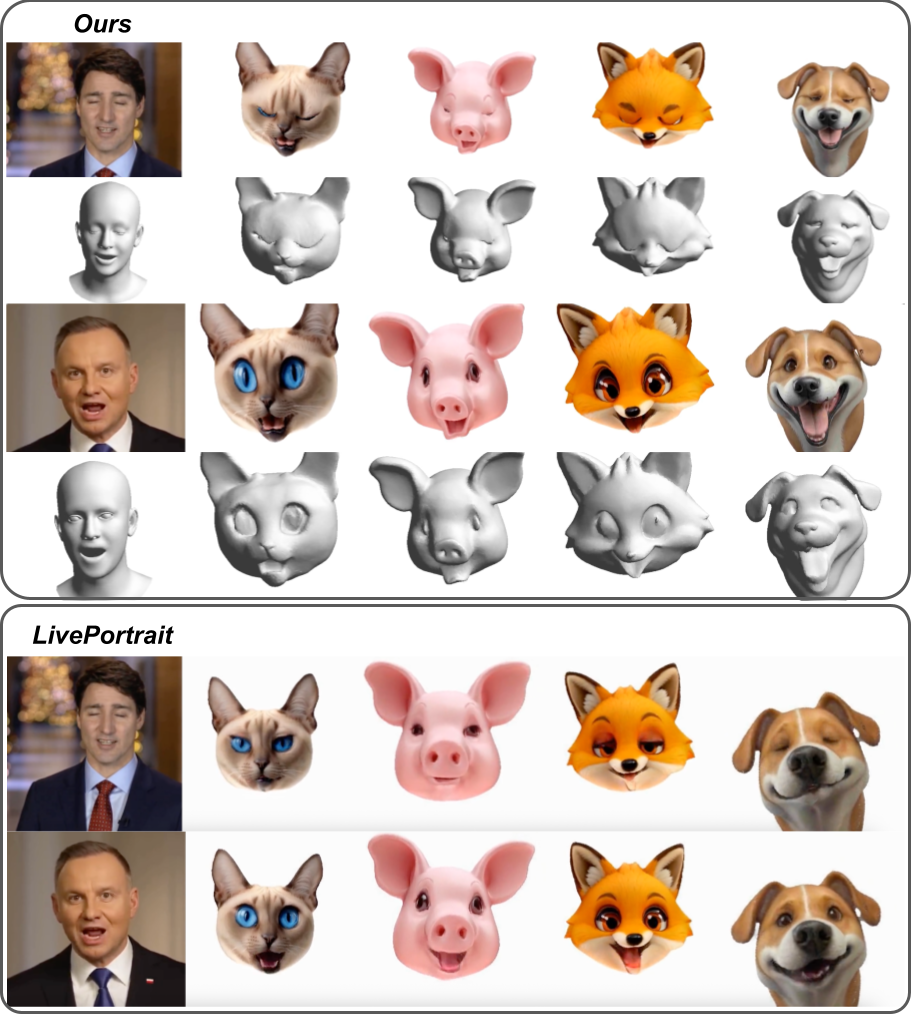}
  \caption{Real-world captures and corresponding retargeted expressions. Top: 3D-aware retargeting by T2Bs, rendered with both textured and geometry only. Bottom: 2D-only retargeting by LivePortrait~\cite{guo2024liveportrait}. T2Bs achieves better retargeting, especially for non-human features such as exaggerated eyes and mouth shapes.}
  \label{sync_human}
  \vspace{-10pt}
\end{figure}

\noindent\textbf{Retargeting.}
Given human face tracking, we update the virtual character’s landmarks based on the relative positions of six landmarks on each eye and eight on the mouth from the human facial landmarks. We then fit the 100 blendshape model by optimizing its weights to minimize a combination of landmark distance $
\mathcal{L}_{\text{lm}}$ and as-rigid-as-possible (ARAP) regularization $\mathcal{L}_{\text{ARAP}}$, with weightings of 1.0 and 0.1. For the eye regions, we introduce additional regularization. 

Fig.~\ref{sync_human} shows examples of the alignment between human and virtual character expressions. We compare with LivePortrait~\cite{guo2024liveportrait}, an image base retargeting method also replies on landmarks. While LivePortrait fails to close large eyes or open wide mouths, our method, achieved by smooth 3D geometry, is more expressive to handle non-human features.

% \vspace{-5pt}
\subsection{Ablation Studies}
\label{ablation}
In this section, we conduct ablation studies on proposed components. We show qualitative comparison with our virtual characters, and show Quantitative comparison on Objaverse-XL~\cite{deitke2023objaverse} samples in Supp. material.

\begin{figure}[t]
  \centering
  \includegraphics[width=\linewidth]{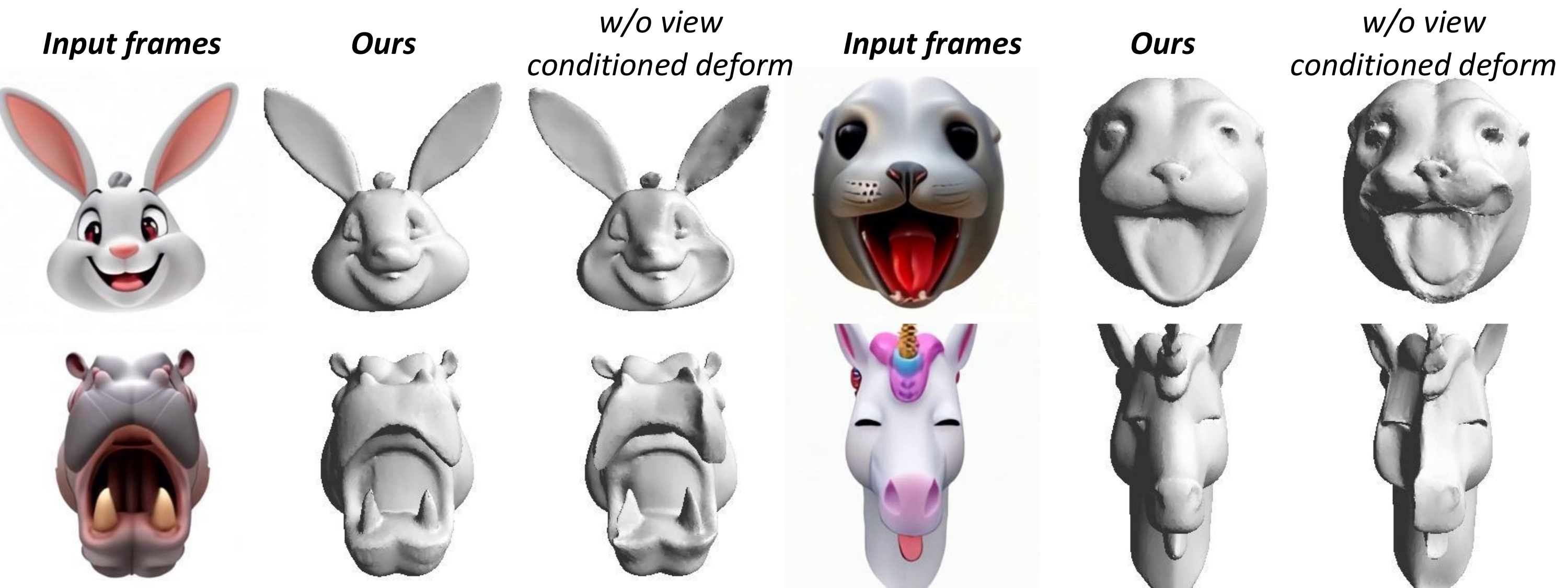}
  \vspace{-15pt}
  \caption{Ablation study on view conditioning: Without view conditioning, the model develops artifacts and fails to maintain coherent geometry due to 3D inconsistencies in the generated 4D videos.}
  \label{fig:ablation_view}
\end{figure}

\noindent\textbf{View Conditioning.}
We introduce view dependency into the deformation represention to address 3D inconsistencies in generated 4D videos. To validate this design, we compare reconstructions without view-conditioned deformation as shown in Fig.~\ref{fig:ablation_view}. For clarity, geometry is visualized as a mesh with the same topology as the static 3D asset. Without view conditioning, multi-view inconsistencies in the generated video frames lead to artifacts and geometric distortions in the reconstructed meshes.

\begin{figure}[t]
  \centering
  \includegraphics[width=\linewidth]{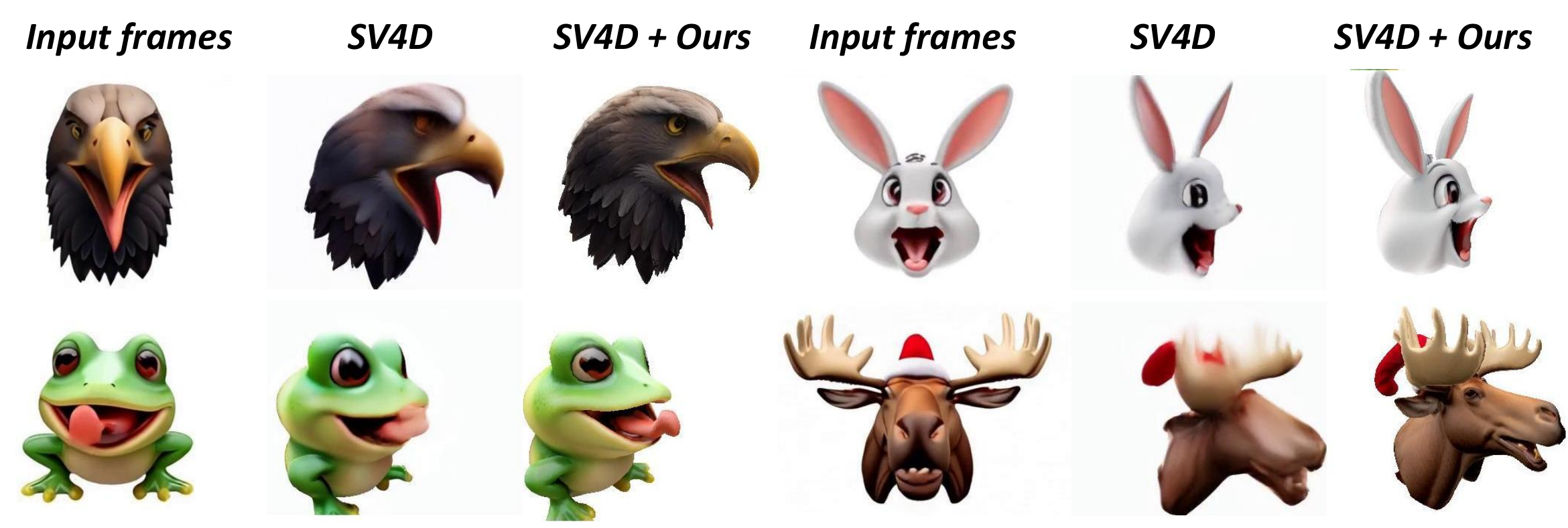}
  \vspace{-15pt}
  \caption{Ablation study of changing 4D video generation method to SV4D. Our method remains robust to SV4D's blurry outputs, producing improved results over its raw output.}
  \label{fig:sv4d}
\end{figure}

%\subsubsection{Multi-view videos}
\noindent\textbf{Source of Multi-view Videos.}
We optimize VCDGS using videos generated by 4Real-Video. To assess whether this specific 4D video generation method is necessary, we conduct an ablation study using 4D videos from other sources. Specifically, we fit VCDGS on SV4D~\cite{sv4d} outputs while keeping all other settings unchanged. As shown in Fig.~\ref{fig:sv4d}, incorporating VCDGS improves SV4D’s results. SV4D alone tends to produce blurry novel views, while our model remains robust to blurry inputs and generates high-quality renderings with well-detailed geometry.

\begin{figure}[t]
  \centering
  \includegraphics[width=\linewidth]{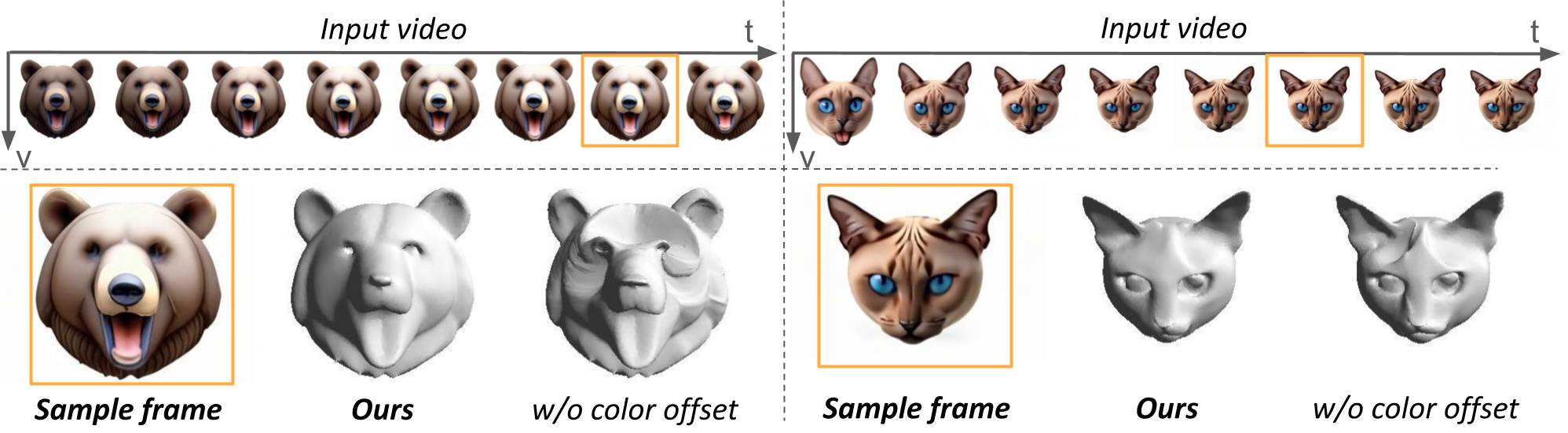}
  \vspace{-15pt}
  \caption{Ablation study on color offset prediction. (Top) Two video examples with noticable variations in color and lighting. (Bottom) Reconstructions with and without the predicted color offset. Without color refinement, shape deformation tends to overfit, leading to suboptimal reconstruction quality.}
  \label{color_offset}
\end{figure}

%\subsubsection{Color offset prediction}
\noindent\textbf{Color Offset Prediction.}
We conducted an ablation study to evaluate the impact of predicting color offsets with RefineMLP. Fig.~\ref{color_offset} (top) provides two examples of common color and lighting inconsistency issues that arise during video generation. When the Gaussian color is fixed, the model struggles to converge in affected areas, leading to overfitting in shape deformation, as shown in Fig.~\ref{color_offset} (bottom). Color offset prediction models the lighting changes, allowing for better geometry reconstruction.

\begin{figure}[t]
  \centering
  \includegraphics[width=\linewidth]{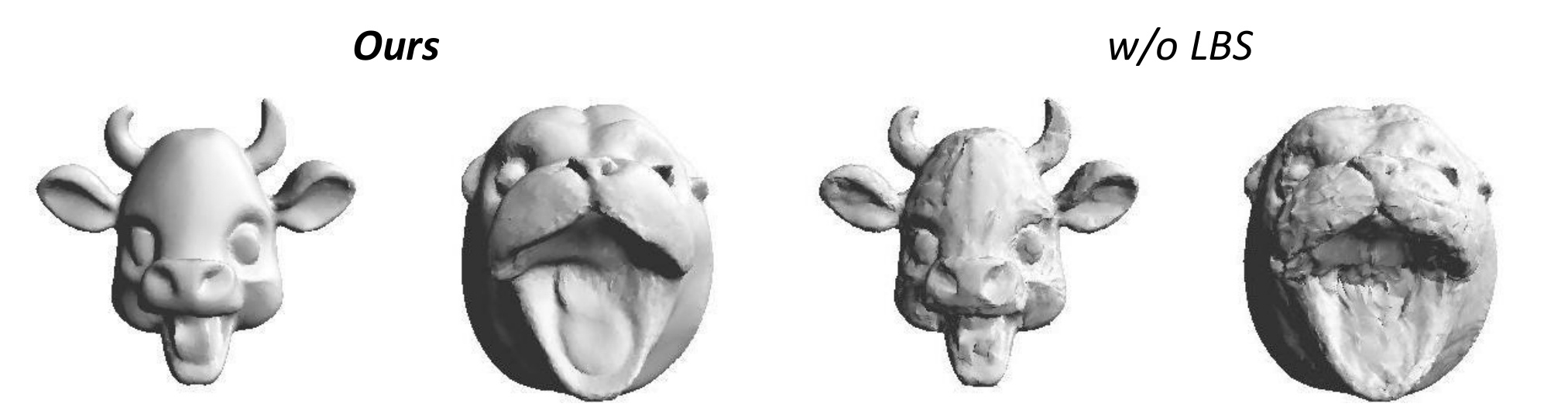}
  \vspace{-15pt}
  \caption{Ablation study on integrating Linear Blend Skinning (LBS) in deformation. Removing LBS leads to degraded reconstruction quality.}
  \label{LBS}
\end{figure}

%\subsubsection{Blending weights prediction}
\noindent\textbf{Linear Blend Skinning.}
LBS employs deformation primitives to ensure well-regularized deformations.
Fig.~\ref{LBS} illustrates the impact of the LBS on geometry reconstruction. We compare our method with a commonly used approach~\cite{4dgs} that directly predicts per-Gaussian deformation. Our method produces smooth, high-quality reconstructions, whereas the comparison results exhibit significant noise.

\section{Conclusion}
We introduced \sname a novel framework for generating high-quality, animatable 3D character blendshape models from textual descriptions. By leveraging advancements in text-to-3D generation and video diffusion models, our pipeline bridges the gap between static geometry and dynamic motion. Extensive evaluations confirm its effectiveness in producing expressive and visually coherent 3D character models.

\noindent\textbf{Limitations.} Our method is designed for head animations and does not extend to full-body motion, as blendshape models are unsuitable for non-linear articulated motions. While it effectively models challenging geometries like protruding tongues, it relies on the initial 3D asset having sufficient coverage of the mouth interior, as 4D reconstruction is sensitive to rapid topological changes. Additionally, the video diffusion model enables diverse expression generation but occasionally produces unnatural deformations or temporal inconsistencies. Future advancements in video generation may help address these limitations.
%Our approach introduces a view-dependent deformation framework using Gaussian splatting to address multi-view inconsistencies.
%Extensive evaluations, including comparisons with state-of-the-art 4D generation methods and ablation studies, demonstrate our method's performance in generating expressive and visually coherent 3D character models. 

%Additionally, the proposed pipeline enables the construction of a robust expression morphable model that generalizes effectively to new expressions, showcasing its versatility and practical applicability in animation and virtual character creation.
%
%This work sets a new promising direction for expressive 3D character modeling and opens avenues for future research, including extending the framework to arbitrary 3D shapes and complex scenes.
\section{Appendix}

\subsection{Expression prompts}

We generate videos using expression prompts concatenated with the original character prompt and camera motion description. To ensure a diverse range of head motions for virtual characters, we predefine a set of expression prompts that encompass various potential movements. These expressions can be categorized into two groups: 

Physically specific expressions – These describe concrete, observable actions involving facial features: \texttt{talking}, \texttt{screaming}, \texttt{laughing}, \texttt{smiling}, \texttt{smirking}, \texttt{closing the mouth}, \texttt{opening the mouth extremely wide}, \texttt{blinking}, \texttt{teasing the eyes}, \texttt{looking around}, \texttt{waving the ears}, \texttt{tongue sticking out the mouth}, \texttt{shaking the head}. 

Emotionally expressive states – These convey the character's inner feelings and overall demeanor: \texttt{sad}, \texttt{angry}, \texttt{chilling}, \texttt{happy}, \texttt{pensive}, \texttt{confused}, \texttt{disappointed}.

We observe that, on one hand, the character may exhibit additional expressions beyond those specified in the input prompt, such as closing the mouth while blinking. On the other hand, the video model may fail to generate the described expression as prompted, especially the emotionally expressive states. In the latter case, we retain the video if it still presents natural-looking motion.

\subsection{T2Bs model Expressiveness Evaluation details}
As demonstrated in Fig. 5 of the main paper, we fit T2B models to random captures that fall outside the model's training range. Specifically, we fit the corresponding models on 10 identities, each with 5 held-out expression videos, and then fit the model to each frame of these videos. The average pixel-wise L2 fitting error is 0.0009, while the average 3D point-to-point error is 0.0017 relative to the bounding box size.

\subsection{Analysis on the Number of Control Points}

\begin{figure}[h]
  \centering
  \includegraphics[width=\linewidth]{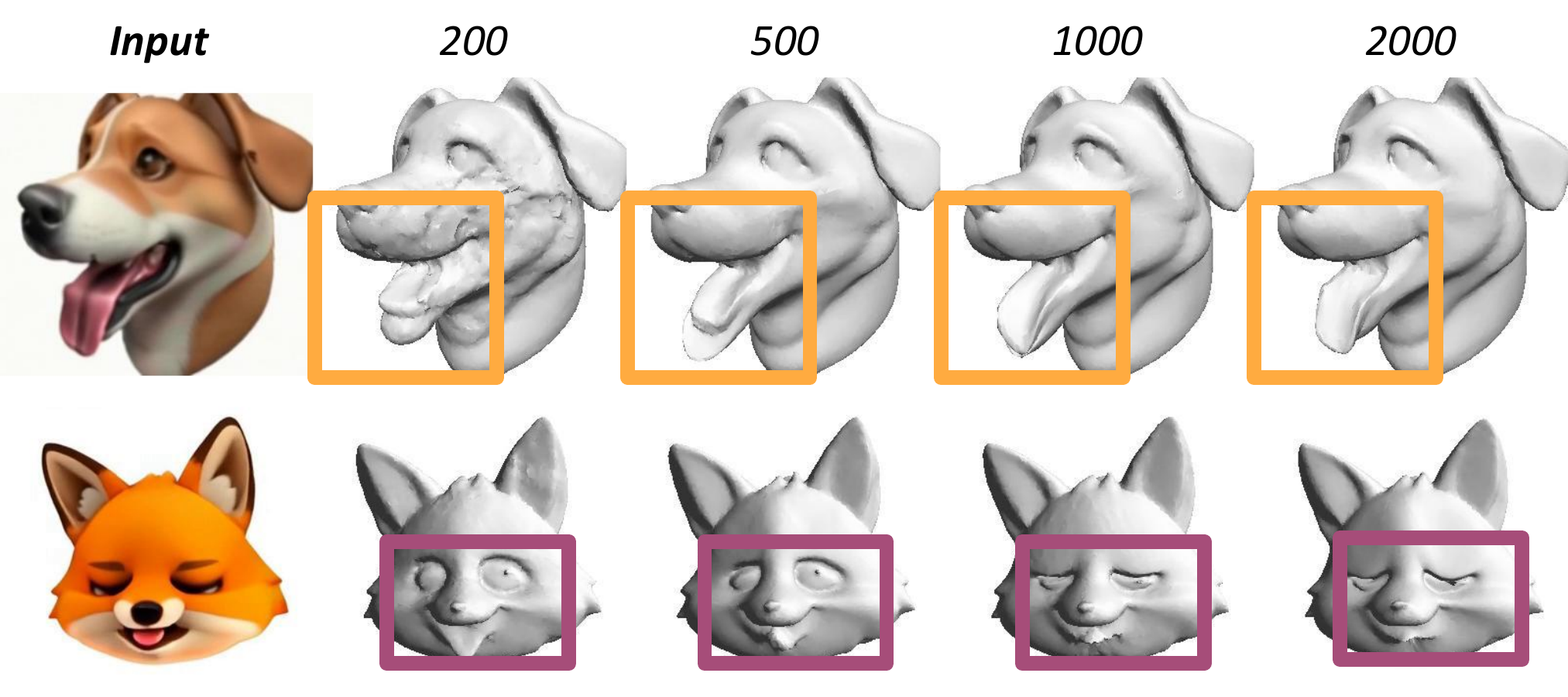}
  \caption{Ablation study on the number of control points. Using 2000 joints captures fine-grained motions, such as tongue (1st row) and eyelid (2nd row) movements, better than 200, 500, or 1000 joints.}
  \label{fig:ablation_num_joints}
\end{figure}

\begin{figure}[t]
\vspace{-10pt}
\centering
\includegraphics[width=\linewidth]{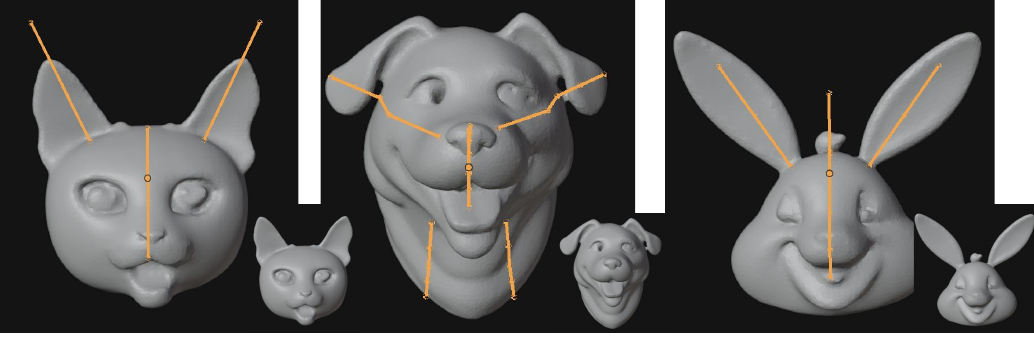}
\vspace{-12pt}
\caption{(Skeleton predicted by UniRig (orange), which isn't suitable for facial motion.}
\label{fig:sds-bone}
\vspace{-15pt}
\end{figure}

To ensure scalability, we use pre-defined control points from the static asset instead of jointly optimizing them across all expression videos. Specifically, we first apply the KNN algorithm to select \( k = 2000 \) uniformly distributed control points. For each Gaussian, we assume it is influenced by its 10 nearest control points. The blending weights for these neighbors are computed as the normalized inverse Mahalanobis distances. We then fix both the Gaussian-to-control-point associations and the blending weights, and optimize only the transformation of each control point. This allowing new expression videos to be incorporated without requiring re-optimization of previously processed videos. Also, this modular approach avoids the need for computationally expensive joint optimization across an ever-growing dataset. We show parameter analysis on the the number of control points in Fig~\ref{fig:ablation_num_joints}.  2000 control points captures fine-grained motions, such as tongue (1st row) and eyelid (2nd row) movements, better than 200, 500, or 1000 joints.

Beyond KNN control points, we also try to obtain skeletons predicted by, MagicArticulate, and UniRig. We show a few examples of bones and control points prediction in Fig.~\ref{fig:sds-bone}, which is not suitable to model facial expressions.

\begin{table*}[t]
\centering

\renewcommand\tabcolsep{5pt}
\resizebox{\textwidth}{!}{
\begin{tabular}{lcc|ccc|ccc}
\toprule
image quality & LPIPS$\downarrow$ & FID$\downarrow$  & geometry & p2p$\downarrow$  & NC$\downarrow$  & geometry & p2p$\downarrow$  & NC$\downarrow$   \\
\midrule
SV4D        & 0.1543 & 167.3 & T2Bs & \textbf{0.0576} & \textbf{0.1611} & T2Bs & \textbf{0.0476}  & \textbf{0.1671} \\
4Real-Video       & 0.1824 & 151.7 & w/o view conditioning & 0.0632 & 0.1713 &  &  &   \\
T2Bs        & \textbf{0.0880}  & 59.6 & w/o RefineMLP & 0.0613 & 0.1851  & w/o RefineMLP  & 0.0537  & 0.2430   \\
T2Bs (SV4D) & 0.0882  & \textbf{56.6}  & \cite{4dgs} & 0.0696 & 0.2654 &   &   &   \\
\bottomrule
\end{tabular}
}
\vspace{-10pt}
\caption{\textbf{(Left)} Quantitative comparison of 4D generation on per-frame image quality when changing the source of multi-view video. T2Bs with the source of both 4Real-Video (T2Bs) and SV4D (T2Bs(SV4D)) archieve high-quality results and improve significantly from the inputs. \textbf{(Middle)} Effect of each proposed component on geometry accuracy and smoothness. \textbf{(Right)} Effect of RefineMLP on appearance inconsistencies \textbf{across time}. Abbreviation: p2p - point to point euclidean distance, NC - normal consistency.}
\label{tab:quantitative}
% \vspace{-10pt}
\end{table*}

\begin{table*}[t]
\centering
\includegraphics[width=\linewidth]{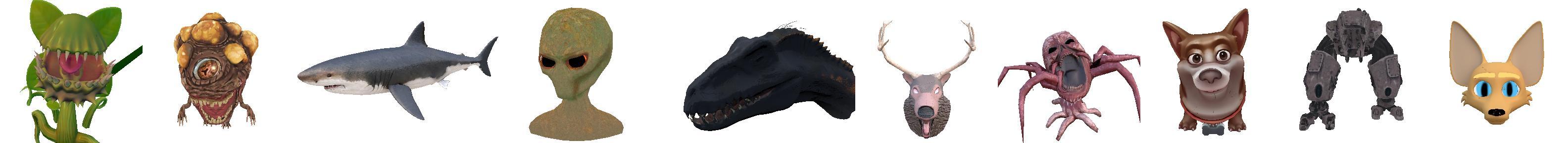}
% \vspace{-10pt}
\renewcommand\tabcolsep{5pt}
\resizebox{\linewidth}{!}{
\begin{tabular}{l|c|c|c|c|c|c|c|c|c|c|c|c}
\toprule
LPIPS & plant & bug & shark & alien & indoraptor & deer & cryinghead & corgi & robot & fox & average \\
\midrule
SV4D        & 0.2405 & 0.2270 & 0.0928 & 0.1356 & 0.1681 & 0.1168 & 0.1441 & 0.1078 & 0.1887 & 0.1219 & 0.1543 \\
4Real-Video       & 0.2096 & 0.1986 & 0.1627 & 0.1872 & 0.1754 & 0.1761 & 0.1797 & 0.1624 & 0.1759 & 0.1960 & 0.1824 \\
T2Bs        & \textbf{0.0938} & \textbf{0.1365} & 0.0610 & 0.1065 & 0.0995 & \textbf{0.0697} & \textbf{0.0562} & 0.0681 & 0.1065 & 0.0817 & \textbf{0.0880} \\
T2Bs (SV4D) & 0.1388 & 0.1734 & \textbf{0.0454} & \textbf{0.0989} & \textbf{0.0910} & 0.0787 & 0.0575 & \textbf{0.0357} & \textbf{0.0943} & \textbf{0.0679} & \textbf{0.0882} \\
\bottomrule
\end{tabular}
}

\resizebox{\linewidth}{!}{
\begin{tabular}{l|c|c|c|c|c|c|c|c|c|c|c|c}
\toprule
FID & plant & bug & shark & alien & indoraptor & deer & cryinghead & corgi & robot & fox & average \\
\midrule
SV4D        & 179.2400 & 293.5054 & 49.7282 & 144.4213 & 141.5446 & 166.7694 & 171.7080 & 28.2255 & 411.4537 & 86.6902 & 167.3286 \\
4Real-Video & 112.5859 & 240.1668 & 69.2304 & 168.9460 & 199.0181 & 214.8099 & 174.2843 & 50.8184 & 191.6526 & 95.3293 & 151.6842 \\
T2Bs        & \textbf{32.5861} & 125.0183 & \textbf{29.5620} & \textbf{69.0403} & \textbf{80.6074} & \textbf{49.4793} & \textbf{34.8517} & 20.5047 & 103.4165 & 51.2718 & 59.6338 \\
T2Bs (SV4D) & 48.3140 & \textbf{94.9072} & 32.0060 & 74.2109 & 101.4071 & 55.2417 & 35.6081 & \textbf{14.8072} & \textbf{65.7210} & \textbf{43.6929} & \textbf{56.5916} \\
\bottomrule
\end{tabular}
}

\resizebox{\linewidth}{!}{
\begin{tabular}{l|c|c|c|c|c|c|c|c|c|c|c|c}
\toprule
p2p & plant & bug & shark & alien & indoraptor & deer & cryinghead & corgi & robot & fox & average \\
\midrule
T2Bs        & \textbf{0.0529} & \textbf{0.1243} & \textbf{0.0603} & \textbf{0.0398} & 0.0400 & \textbf{0.1056} & \textbf{0.0174} & \textbf{0.0189} & \textbf{0.0439} & \textbf{0.0731} & \textbf{0.0576} \\
{\cite{4dgs}}      & 0.0543 & 0.1553 & 0.0755 & 0.0550 & \textbf{0.0365} & 0.1173 & 0.0230 & 0.0259 & 0.0577 & 0.0951 & 0.0696 \\
\bottomrule
\end{tabular}
}

\resizebox{\linewidth}{!}{
\begin{tabular}{l|c|c|c|c|c|c|c|c|c|c|c|c}
\toprule
NC & plant & bug & shark & alien & indoraptor & deer & cryinghead & corgi & robot & fox & average \\
\midrule
T2Bs        & \textbf{0.0874} & \textbf{0.2207} & \textbf{0.2070} & \textbf{0.0745} & \textbf{0.2087} & \textbf{0.0715} & \textbf{0.2465} & \textbf{0.1450} & \textbf{0.1606} & \textbf{0.1895} & \textbf{0.1611} \\
{\cite{4dgs}}      & 0.1593 & 0.3805 & 0.2952 & 0.1156 & 0.3145 & 0.3562 & 0.3486 & 0.1790 & 0.1917 & 0.3129 & 0.2654 \\
\bottomrule
\end{tabular}
}

\vspace{-7pt}
\caption{(Top) Each sample we use from Objaverse-XL dataset. (Botton) Per-identity improvement in LPIPS, FID and geometry improvement compared to \cite{4dgs}. Abbreviation: p2p - point to point euclidean distance, NC - normal consistency.}
\label{tab:quantitative-sample}
\vspace{-5pt}
\end{table*}

\subsection{Ablation studies on Objaverse-XL samples}

We further evaluate our method on 10 artist-animated virtual characters that is closely aligned with our application domain from Objaverse-XL. For each identity, we baked geometry sequences with a shared texture. We rendered (ambient=1.0) fixed-time, fixed-view videos as the pipeline input, and full sequences across all times and views as ground truth.

Table~\ref{tab:quantitative} (Left) shows 4D generation quality when switching source multi-view videos from 4Real-Video (T2Bs) to SV4D (T2Bs), which corresponds to Fig. 7. Table~\ref{tab:quantitative} (Left) shows the effect of view conditioning, RefineMLP and integrating Linear Blend Skinning (LBS), which correspond to Fig. 6, 8, 9.

Specifically, as for RefineMLP, in order to solve the concern that RefineMLP might be exploited to compensate for geometry, but with GT geometry, we show RefineMLP actually improves the geometry in Tab. \ref{tab:quantitative} (Middle). We attribute this to RefineMLP effectively handling \textbf{appearance inconsistencies across views} from 4D generation, since we render input fixed-view videos without appearance inconsistency. To further evaluate RefineMLP, we simulate \textbf{appearance inconsistencies across time} by multiplying extreme noise $\mathcal{U}(0.5, 1.5)$ to the texture map, clamped to $[0, 1]$. Instead of running 4D generation, we render the geometry sequence with different noisy textures, ensuring there is no inconsistency across views. As shown in Tab. \ref{tab:quantitative} (Right), RefineMLP still improves geometry quality.

\begin{figure*}[t]
  \centering
  \includegraphics[width=\textwidth]{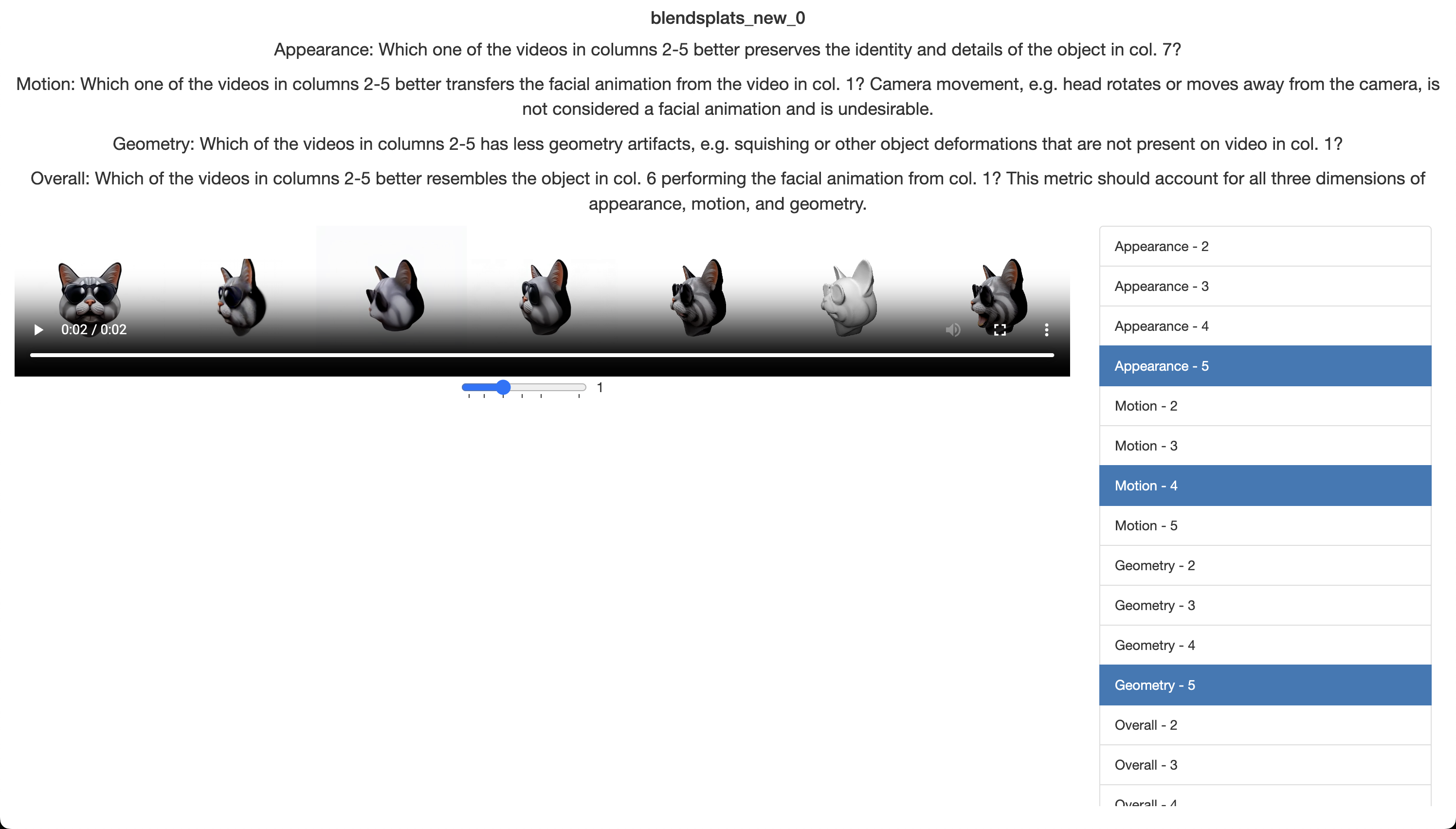}
  \caption{A screenshot of the user study interface. Participants are presented with a set of four single-choice questions, each designed to identify the best-performing column for a given dimension. The data is randomly shuffled to mitigate any potential ordering bias.}
\label{fig:user_study_interface)}
\end{figure*}

The data of animatable 3D animal head model is limited even in Objaverse-XL. We further shows the per-identity comparison in 4D generation and geometry in Table~\ref{tab:quantitative-sample}. It's clear to see that T2Bs improve significantly on \textbf{each} identity.

\subsection{User Study Interface}
We demonstrate the user interface of our user study in Fig~\ref{fig:user_study_interface)}. We provide participants with video comparisons of VCDGS and baseline methods. They are free to replay the videos until they make their judgments, selecting the four best-performing columns based on four different criteria. Participants can also use the provided slider to zoom in and out, especially to zoom in for detailed appearance differences.

        % uncommend it if Arxiv

{
    \small
    \bibliographystyle{ieeenat_fullname}
    \bibliography{main}
}

\end{document}